# USING 2.4 GHZ WIRELESS BOTNETS TO IMPLEMENT DENIAL-OF-SERVICE ATTACKS

*Buryachok V. L.*, DSc,
*Sokolov V. Yu.*, MSc

*Ukraine, Kyiv, State University of Telecommunications, Dept. of Information and Cyber Security*

**ABSTRACT**

This article attempts to create a software and hardware complex that can work autonomously and demonstrates the ease of implementation of attacks on denial of service on wireless networks, which in turn emphasizes the need to provide comprehensive protection of wireless networks.

**Introduction.** Wireless networks have become very widespread because their configuration and usage is very convenient for ordinary users. However, unlike wired networks, wireless is secured to me, since intercepting information in them does not require physical connection to the network.

The object of research in the work is the robustness of the wireless network during an attack on denial of service. The subject of the study is an attack using a wireless botnet. The purpose of the work is to effectively carry out an attack on wireless networks IEEE 802.11 with the help of a wireless botnet. As a result, the task of constructing the architecture and its implementation as a distributed bots network, which can be used to demonstrate an attack failure on service to wireless access points (APs), is solved.

In the work the example of the hardware-software implementation of the wireless botnet and the experimental stand for the wireless network is given. Possible areas for developing this work are related to the study of methods for protecting wireless APs.

In the wired network packets, the data is transmitted in a physical environment such as copper cable or fiber optics. In the wireless system, your data is transmitted literally around the air around you. In addition, physical access is not required to access the network. Because of this, malicious people have far more opportunities to inflict damage on network infrastructure and remain unnoticed.

Wireless attacks can be carried out in a variety of ways. Some methods are aimed at cheating users, others use brute force, and some are looking for people who do not bother themselves in protecting their network. Many of these attacks are combined with each other in real use. Let's look at examples of such attacks [1–3].

Packet sniffing—because wireless traffic is transmitted in an unprotected environment, it is easy to capture. Quite a lot of traffic (FTP, HTTP, SNMP, etc.) is sent in the open, that is, without encryption, and the data can be received as plain text. Thus, using a tool such as Wireshark, anyone can read your data in plain text. This can lead to the theft of passwords or leakage of confidential information. Encrypted data can also be captured, but it's much more difficult for an attacker to decrypt such data packets.

Rogue AP—when an unauthorized AP appears in the network, it can be considered fraudulent. It may appear due to an employee's mistake or with an attacker. Such APs can become vulnerable to the network because it may not be protected from various attacks, which include vulnerability scans in preparation for an attack, ARP requests, packet capture, and denial-of-service attacks.

Password theft—when you exchange data over a wireless network, you often log in to websites. You send passwords over the network, and if the site does not use SSL or TLS, this password is transmitted as text, and the attacker can read it. In fact, even when sending encrypted traffic, there are ways to bypass these encryption methods to steal your password.

Man-in-the-middle—malicious people can deceive communication devices by sending their data to their system. Here they can record traffic for further viewing (for example, when intercepting packets) and even change the contents of files. Inside these packages, different types of malware can be placed, email content may be modified, or traffic may be deleted to block the connection.

Jamming—such an attack can be carried out in two different ways: the first involves "bombarding" the operating frequency of the attacked host, with different noise traffic, which is "information garbage" and leads to denial of service; the second is to use special noise generators that interfere with data transfer at the physical level.

Availability is one of the three core concepts of computer security, along with confidentiality and integrity. Nevertheless, apart from, in fact, the availability of information, the most important role is played by the ability to get and use it in a timely manner. Based on this, accessibility can be defined as the ability to use the necessary information or resource in a reliable and timely manner.

Denial of service is a threat that potentially violates the availability of the resource in the system. In turn, a denial of service attack is an action (or set of actions) performed by the attacker to make the resource inaccessible to its potential users.

In the case of denial-of-service attacks, malicious people may launch their attacks from one or more of the hosts they control. When attacker messages are sent from multiple hosts distributed on the network, this is called a Distributed Denial of Service (DDoS) attack. Sometimes, single attacks are referred to as single-source denial-of-service attacks (SDoS).

Table I shows examples of DoS attacks, depending on the level of the OSI model on which the attack is carried out, as well as the possible consequences of such attacks.

Table 1. Examples and Objectives of Denial-of-Service Attacks According to OSI Levels

| OSI level | Task level | Protocols | Examples of DoS technologies | Consequences of DoS Attacks |
|---|---|---|---|---|
| 7 | Start creating data packets. Joining and accessing data. User-defined protocols such as FTP, SMTP, Telnet | FTP, HTTP, POP3, SMTP | HTTP GET and POST requests (website forms: login, photo/video upload, confirmation feedback) | Lack of resources. Excessive consumption of system resources by services on the attacked server |
| 6 | Broadcast data from sender to recipient | Compression protocols, data coding (ASCII, EBCDIC) | Counterfeit SSL queries: Encrypting SSL packets requires a lot of resources, hackers use SSL for HTTP attacks on the victim's server | The attacked systems may stop accepting an SSL connection or automatically reboot |
| 5 | Manage installation and termination of connection, synchronization of sessions within the OS through the network | Input/output protocols (RPC, PAP) | The Telnet attack uses the weak points of the Telnet server software on the switch to make the server unavailable. | Disables the administrator to control the switch |
| 4 | Providing information transfer between nodes without errors, managing the transmission of messages at 1, 2, and 3 levels | TCP, UDP | SYN-flood, an attack on ICMP requests with altered addresses | Achievement of the limit on the width of the channel or the number of permissible connections, violation of the operation of network equipment |
| 3 | Routing and transferring information between different networks | IP, ICMP, ARP, RIP | ICMP-flood | Reducing bandwidth is attacked by the network and the potential overload of the firewall |
| 2 | Installation and support of communication on the physical level | 802.3, 802.5 | MAC-flood—overflow packets of network switches data | The data streams from the sender to the recipient block the operation of all ports |
| 1 | Data transfer in the physical environment | 100BaseT, 1000 Base-X, 802.3, 802.5 | Mute (for wireless networks) | Failure to transmit any messages |

Recently, wireless technologies and networks have come to the fore, mainly because their configuration and use is very simple for ordinary people. As these networks gain popularity, security and reliability become a critical issue. It is worth paying special attention to IEEE 802.11-based wireless networks, which are a family of physical and MAC protocols, and are used in many APs, ad hoc networks and devices of the IoT (Internet of Things) group. The work of the physical, data link and network layer of the OSI model is implemented in such networks as opposed to wired networks and described in the standard IEEE 802.11. Accordingly, wireless networks have their own weaknesses, and hackers have separate technologies aimed specifically at them.

The general nature wireless environment allows attackers to easily monitor communication between connected devices and run common DoS or DDoS attacks on wireless networks by jamming or interference with communication. Such attacks on the physical level cannot be stopped using conventional security mechanisms. An attacker can simply ignore the access protocol to the media and continuously broadcast its data to the physical environment. Therefore, an attacker either does not allow customers to use wireless, or the input generates its traffic, causing a violation in the network.

At the channel level, attackers can use the vulnerabilities of messages and procedures of the MAC (Media Access Control) protocol. For example, they may falsify deactivating packages or de-association to break the connections between nodes and APs, or send RTS and CTS (Request to Send / Clear to Send) packets with fake duration to pause the transmission of neighboring nodes. Attackers may not follow the deferral procedure, and therefore they always get the chance to first send RTS immediately after the last transfer. In addition, it is possible to exhaust the bandwidth of the network by sending large packets of spam without violating the MAC protocol.

DoS attacks on the network layer are mainly aimed at using routing protocols and forwarding on wireless networks. In particular, ad hoc and IoT networks are vulnerable to these attacks. In addition, network attacks DoS in them are very different from attacks on the Internet. Since routers in wireless networks are computers that may be compromised as end nodes, DoS network attacks on wireless networks can be started from any computer on the network. In addition, the methods of protection against DoS on the Internet, requiring the interaction of routers, in local networks will not work.

**Countering attacks on denial of service.** Despite the tremendous efforts of researchers and experts spent on finding solutions to DoS attacks, they still remain an unresolved issue. There are various technical and non-technical issues that need to be well understood in order to design solutions that fundamentally solve this problem, while ensuring the practical deployment.

The task of preventing DDoS attacks is to detect malicious traffic, since traffic is often legitimate as defined by the protocol. Therefore, there is no direct approach or method for filtering or blocking malicious traffic. In addition, you need to understand the difference between bulk and attacking traffic at the application level.

Bulk attacks use a large amount of information that seeks to suppress the target. This traffic may be specific to the application, but more often it is just random traffic that is sent at high intensity to overuse the available victim resources. Bulk attacks usually use bot networks to increase attack. An example of such attacks may be SYN-flood.

Application-level attacks use certain programs or services in the target system. Usually they bombard the protocol and the port that uses a specific service to make the service unprofitable. Often, these attacks target general services and ports, such as HTTP (TCP, 80th port) or DNS (UDP, 53rd port).

The basic principle of protection against DoS and DDoS attacks is that the protection must be comprehensive. For example, you can try to counteract such attacks separately at each level of the network model OSI:

– *Application level*. Application monitoring is a systematic monitoring of software that uses a certain set of algorithms, technologies and approaches (depending on the platform on which it is used) to detect application vulnerabilities. By identifying such attacks, they can be stopped and monitored forever. At this level, this is accomplished most simply.

– *Presentation level*. For harm reduction, consider features such as the allocation of encryption SSL infrastructure (that is, SSL placement on a separate server, if possible) and testing application traffic for attacks or violation of application-level security policies. A good platform ensures that traffic is encrypted and sent back to the original infrastructure with decoded content.

– *Session level*. Maintaining the network equipment software up to date to reduce the risk of a threat.

– *Transport level*. DDoS traffic filtering, known as blackholing, is a method often used by providers to protect customers from threats such as slowing down network equipment and disclaiming services.

– *Network layer*. Limit the number of processed requests to ICMP (Internet Control Message Protocol) and reduce the possible impact of this traffic on the speed of the firewall and the bandwidth of the Internet channel.

– *Data link level.* Many modern switches can be configured in such a way that the number of MAC addresses is limited to trusted, which pass authentication, authorization, and server logging, and then are filtered.

– *Physical level.* Carry out systematic monitoring of physical network equipment.

**The principles of organization of botnets.** The use of thousands of bots for DDoS organization attacks on corporate and government Internet resources is a very widespread and dangerous phenomenon. To create an army of zombie Internet hosts, intruders usually infect remotely-controlled trojans with machines of ordinary people who have fast Internet access, networks maintained by universities and small businesses. Owners of these machines are typically users with a relatively low level of information security awareness and limited resources to protect their Internet infrastructure.

One of the properties of the Internet is its limited resources, and attackers have traditionally used this by exhausting computing capabilities of computers and networks by flooding their numerous requests, thereby depriving users of access to their services. The model of denial-of-service attacks evolved from one attacking machine to another to several against one. Later, the DDoS attack model has changed, and the attackers began to use several handlers to control and manage a large number of hosts against one goal [4].

In Fig. 1 shows a typical botnet arrangement scheme. First, the attacker finds vulnerable sites on the Internet and expands on them the means of attack—agents. The machines on which the agents are called are called bots. These bots run a hidden channel for communication with the command and control server - the handler that controls the attacker. This communication is usually implemented through IRC (Internet Relay Chat), encrypted channels, bot-specific peer-to-peer networks and even Twitter. After this, the attacker spreads the attacking bots commander, instructing the bots about who, when and how to attack. Starting from the given time of attack, the bots generate attacking traffic to commit an attack.

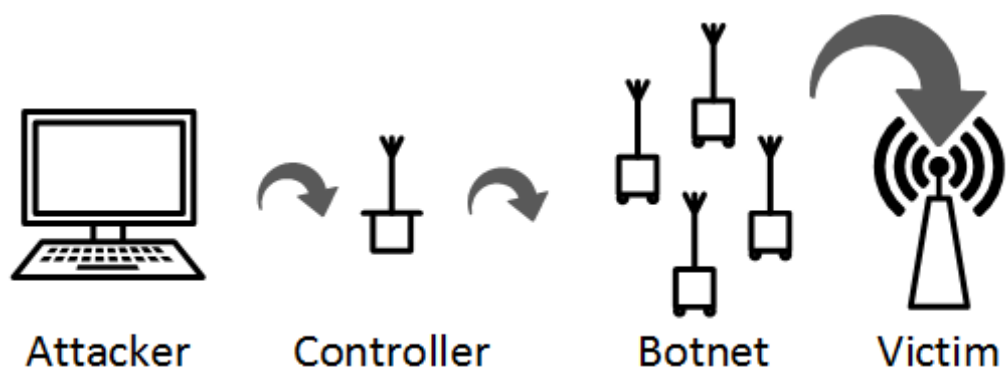

*Fig. 1. A typical botnet structure scheme*

With the advent of cloud services and providers, a new mechanism has appeared. Hackers either lease or compromise large data centers or cloud machines to launch DDoS attacks. Cloud computing not only creates new opportunities for organizations; it also provides an excellent platform for cybercriminals, because it's inexpensive and convenient to use powerful computing resources [5].

**Description of used software components.** The software part of this work uses the ESP8266 SDK and the Arduino API for ESP8266, the main tools for software development for ESP8266, the SDK provides a large selection of tools, and the Arduino API is a convenient and safe use of these tools.

ESP8266 SDK (Software Development Kits)—a suite of IoT application development tools developed by Espressif Systems includes a library of basic features for interacting with hardware and software ESP8266 and examples of projects that can be implemented using them.

Depending on whether they are based on the operating system or not, the SDK can be classified into two types: the Non-OS SDK and the RTOS SDK.

Non-OS SDK is not based on an operating system, it supports AT commands. Uses timers and collages as the main tool for performing various functions—nested events, functions that are caused by certain conditions. Uses the espconn network interface; users need to develop their programs in accordance with the rules for using the espconn interface.

The RTOS SDK is based on FreeRTOS and has open source software on Github. FreeRTOS SDK is based on FreeRTOS, a multitasking OS. You can use standard interfaces to implement resource management, execution delays, interprocess transfer interaction and synchronization, and other solutions that are targeted at performing specific tasks. The RTOS SDK provides a package that provides an interface to the BSD Socket API. Users can directly use the Socket API to develop software applications; Transfer applications from other platforms that use the Socket API to ESP8266, reducing the training costs incurred by changing the platform. The RTOS SDK includes the cJSON library, which simplifies the processing of JSON packets. RTOS is compatible with non-OS SDK in Wi-Fi interfaces and system interfaces, but does not support AT commands.

The Arduino API for ESP8266 was developed on the basis of the ESP8266 SDK, using name consent and the overall functional concept of the Arduino library.

This API consists of several libraries, each of which is designed to simplify the development of software for ESP8266, logically combines low-level SDK features.

Thus, the Arduino API for ESP8266 greatly simplifies and accelerates the development of software for the ESP8266, combining low-level actions.

**Subjects of the study.** Botnet will have the structure of a typical botnet, as in Fig. 1, with one exception—in our botnet there will be only one handler that will interact with all the bots.

Subjects to be involved in the implementation of a denial of service attack:

The processor is the main element of the botnet. It will scan available APs and give the user (administrator) the ability to select the network to be attacked. It will also collect all the information needed for the attack to access the AP and send it to the bots.

The bot is the device that will carry out the attack. After receiving information from the handler, he must prepare the deactivating frames and start sending them, as well as the fake Beacon frames, so that when trying to re-connect, the client did not have the correct AP information.

Administrator is the person who manages the processor. Admin role—select the network to be attacked.

The AP in the work of the botnet does not accept participation.

Client is a device connected to an AP. Will receive from the bots counterfeit frames deauthentication on behalf of the AP [5–10].

**The algorithm of the botnet operation.** In general, the algorithm of the work consists of the following steps:

− Finding AP s in the reach of the handler.
− Select the AP to be attacked.
− Detect devices connected to the selected AP.
− Transmission of information about the AP and its clients, as well as additional information to the bots.
− Formation of bots of frames de-authentication.
− Beacon deactivating packets and fake frames cyclic submit.

**Hardware preparation.** ESP-8266 ESP-01 from Espressif Systems was chosen as the hardware platform for bots (see Fig. 2).

ESP8266 ESP-01—a Wi-Fi module used in projects requiring high-speed wireless transmission between different project objects over Wi-Fi, for example, between a controller and a sensor located at a distance or in an inaccessible location, etc.

Can be used in security systems, remote control systems, home automation systems, telemetry systems.

ESP-01 is equipped with PCB antenna, the reception / transmission range can reach 400 meters.

Therefore, as the voltage required for the module is 3.3V, if the voltage is applied above, the module will fail. Therefore, to use the module, you must first connect the linear power supply AMS1117 to 3.3V (see Fig. 3). The connection scheme of ESP-01 to AMS1117 is shown in Table 2.

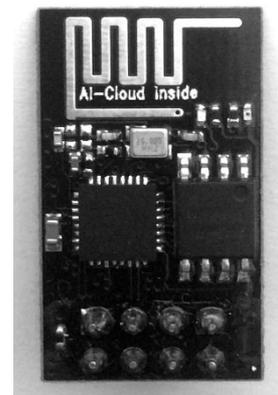

*Fig. 2.* ESP8266 ESP-01

NodeMCU was selected as the handler. NodeMCU—the platform for creating IoT devices based on the Wi-Fi module ESP8266 (ESP-12E). NodeMCU (see Fig. 4) has integrated GPIO, PWM, IIC, 1-Wire, ADC interfaces. There is a USB-UART converter that allows you to program the card using Arduino IDE or Lua.

Table 2. Connection scheme ESP-01 to AMS1117

| ESP-01 | AMS1117 |
|--------|---------|
| GND    | GND     |
| CH_PD  | $V_{out}$ |
| $V_{cc}$ | $V_{out}$ |

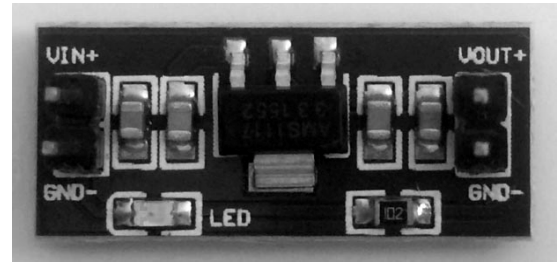

*Fig. 3. AMS1117 to 3.3V*

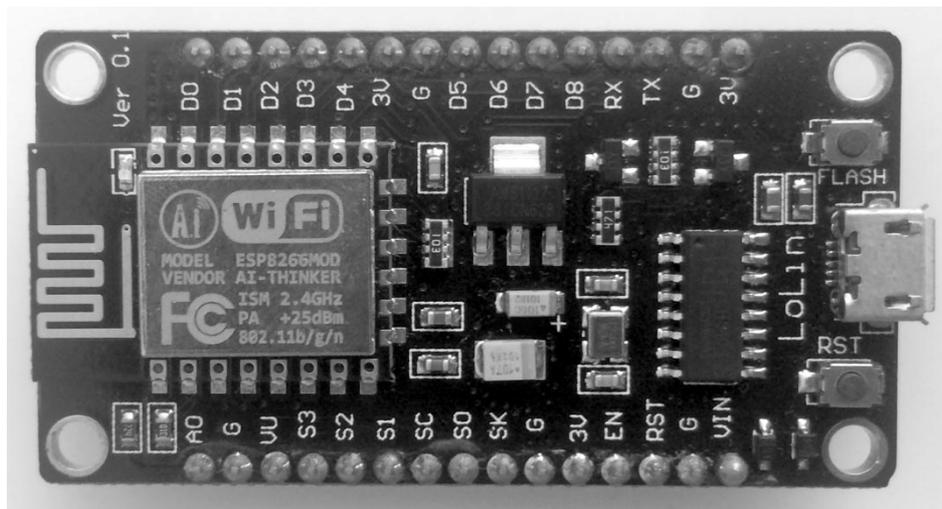

*Fig. 4. NodeMCU*

**Preparing for an attack.** In order to start the attack, you must connect the handler and the bots to the power supply. It should be noted that it is possible to include them in advance, boots after the inclusion will expect a connection with the handler.

From the beginning, the handler creates an AP to which the administrator and the bots connect, which will then receive the task from the handler. After that, the handler launches a simple web server, using which administrator will be able to select an AP to attack. For each request that needs to be processed, you need to set a separate function.

The administrator can connect to the AP created by the handler (by default—esp_ap) and go to the browser on 192.168.4.1. As a result, the administrator will see a page similar in Fig. 5

**Identify AP clients.** In order to find AP clients, we need to move the handler to monitor mode. This can be done using the wifi_promiscuous_enable function, and although it says in the title that it activates the decrypting mode; it will activate the monitor mode. In addition, before activating this mode, we need to specify the function to be called when the handler captures the data packet. She will check the received package and add the recipient to the client list only if the package meets the following conditions: it is sent by the AP, it is sent not to broadcast and not to multicast.

When the crawl is complete, we will get a client list in clients_list and everything will be ready to send the bots to the task.

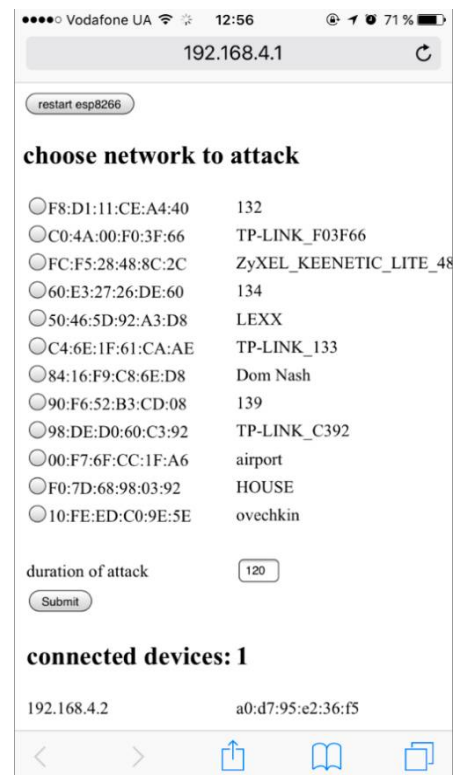

*Fig. 5. Page to select the access point to be attacked*

**Transfer task to bots.** After the handler has formed a list of AP clients, he needs to send the bots all the information they need to start the attack. This information is sent using UDP, as this is the easiest way to transfer plain text between devices in the same network. To create a packet with information about the attack, we will use the template shown in Fig. 6

| channel len | channel | time len | time | clients num len | clients num | SSID len | SSID | MAC AP | MAC 1 | … |
|---|---|---|---|---|---|---|---|---|---|---|
| | | | | | | | | | | |

*Fig. 6. The structure of the UDP sent to the bots*

List of fields:
− channel len—number of characters in the channel.
− channel—the number of the channel on which the AP operates.
− time len—number of characters in time.
− time—the duration of the attack in seconds.
− clients num len—number of characters in clients number.
− clients num—number of clients.
− SSID len—the number of characters in the SSID.
− SSID is the network name.
− MAC AP—MAC address of the AP.
− MAC—MAC address of the client.

**Formation of deactivating frames and start of attack.** All this time bots were waiting until the handler sent them a packet with the information necessary to carry out the attack. After receiving this package, each bot disassembles it and generates the same access list client that was in the handler. Further, on the basis of this list, each bot generates a new one, which does not already have MAC addresses stored, but the deactivating frames that will be sent to point clients on behalf of the AP. Deauthentication frames are formed by the deauth_frame function.

Also, behind the scenes of deauthentication, bots send fake Beacon frames generated using the AP SSID, and the rest of the information is generated accidentally. To create a Beacon frame, the sendBeacon function is used.

Therefore, if the user of the disassociated device tries to reconnect to the AP again, it will have incorrect information about it, so it will not be able to connect again.

**Test bench.** To test the botnet's performance, a test bench was created, also based on NodeMCU with OLED display SSD1306.

This bench can work as a repeater to build or test the stability of IOT wireless systems. It connects to an AP whose parameters are predefined. If there is no connection, the test stand simply displays the devices connected to it in real time. All information is displayed on the connected OLED display, as shown in Fig. 7.

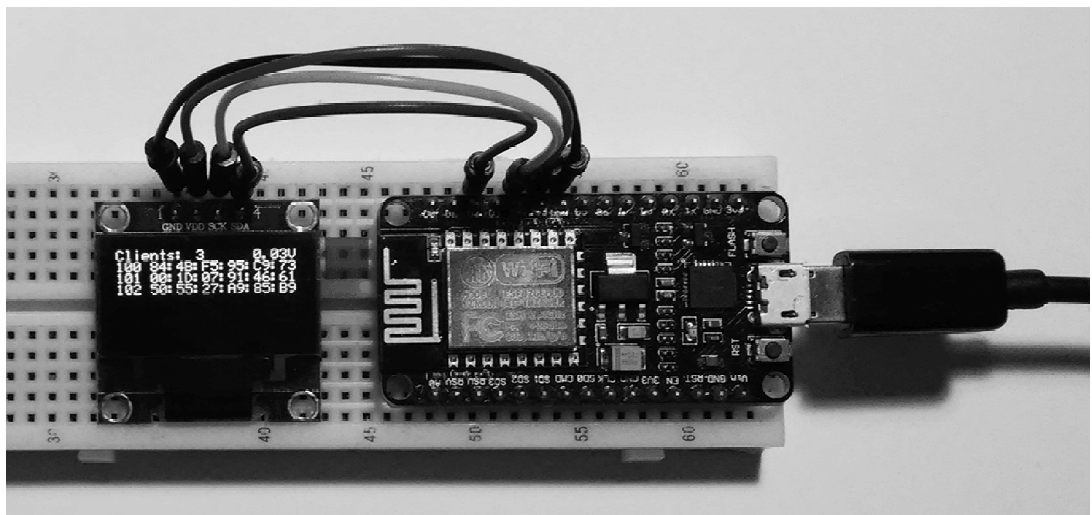

*Fig. 7. Test bench*

**Conclusions.** In today's realities it is very important to pay special attention to computer security regardless of whether you are the administrator of a large corporate network or a simple computer user. Technologies aimed at the theft of confidential data or violation of the normal operation of various services are developing faster than the means of their protection; malicious people do not obey any rules, do not adhere to standards, they achieve their goals in any way. The switch from solitary abandoned service to distributed distributions is a great proof: hackers capture other people's devices and use them for their own purposes without the consent of the owners. Therefore, to defend itself from such an enemy to follow the basic rules of protecting computer systems is not enough, it is necessary to provide a comprehensive protection of their system.

This paper was considered one of the main security threats on the Internet—denial of service were given a comprehensive overview and classification of DoS-attacks and countermeasures types and is designed to implement wireless botnet attacks to denial of service.

Particular attention is paid to wish that the developers ESP8266 SDK, on which is built the botnet is now understood that the ability to send any personnel management can be dangerous, and can be used to do harm. That is why the new versions of the ESP8266 SDK no longer have this capability.

After many tests, one can definitely say that the system is working properly. An important point is that the system can be scaled up by increasing the number of bots or by reducing, leaving only one. If one or more bots are outside the reach of the handler during the attack, it will not affect the work of other bots and the handler (they will wait for the connection). In addition, it should be noted that during the attack, you can change the AP to another, starting a new attack.

However, when searching for AP clients, some devices may remain unlabeled if they do not actively exchange data from the AP at the time of scanning, for example, if it is a smartphone in standby mode. Tests have shown that the more active the data exchange with the device with the device, the greater the likelihood that the device will be disconnected from the AP.